\documentclass[journal]{IEEEtran}

\ifCLASSINFOpdf
\else
   \usepackage[dvips]{graphicx}
\fi
\usepackage{url}

\usepackage{graphicx}
\usepackage{amsmath,amsfonts}
\newtheorem{remark}{Remark}

\begin{document}

\title{Cubature Kalman filter Based on generalized minimum error entropy with fiducial point}

\author{Jiacheng He, Gang Wang, Zhenyu Feng, Shan Zhong, Bei Peng
\thanks{The NNSFC funded this research with Grant 51975107, together with the Sichuan Science and Technology Major Project Nos. 2022ZDZX0039, No. 2019ZDZX0020, and No. 2022YFG0343. (Corresponding author: Bei Peng.)}
\thanks{J. He, Z. Feng, S. Zhong, and B. Peng are with the School of Mechanical and Electrical Engineering, University of Electronic Science and Technology of China (UESTC) (e-mail: hejiacheng\_123@163.com; zhenyu.feng.uestc@gmail.com; 2608589754@qq.com; beipeng@uestc.edu.cn).}
\thanks{G. Wang is with the School of Information and Communication Engineering, UESTC (e-mail: wanggang\_hld@uestc.edu.cn).}}

\markboth{GGGGG, GGGGGG, GGGG, GGG}
{Shell \MakeLowercase{\textit{et al.}}: Bare Demo of IEEEtran.cls for IEEE Journals}
\maketitle

\begin{abstract}
In real applications, non-Gaussian distributions are frequently caused by outliers and impulsive disturbances, and these will impair the performance of the classical cubature Kalman filter (CKF) algorithm. In this letter, a modified generalized minimum error entropy criterion with fiducial point (GMEEFP) is studied to ensure that the error comes together to around zero, and a new CKF algorithm based on the GMEEFP criterion, called GMEEFP-CKF algorithm, is developed. To demonstrate the practicality of the GMEEFP-CKF algorithm, several simulations are performed, and it is demonstrated that the proposed GMEEFP-CKF algorithm outperforms the existing CKF algorithms with impulse noise.
\end{abstract}

\begin{IEEEkeywords}
cubature Kalman filter, GMEEFP, impulse noise.
\end{IEEEkeywords}

\IEEEpeerreviewmaketitle

\section{Introduction}

\IEEEPARstart{F}{or} linear dynamic systems influenced by white Gaussian noise, the Kalman filter offers the best solution for state estimation problems utilizing the minimum mean square error criterion. Numerous nonlinear extensions, including extended KF (EKF) \cite{bar2001estimation}, unscented KF (UKF) \cite{847726}, cubature KF (CKF) \cite{4982682}, and their variants, have been derived for nonlinear dynamical systems. The CKF is widely applied as a result of its third-order computational accuracy and greater numerical stability \cite{6746072}. In reality, non-Gaussian noise \cite{9340372,8876690,8398426,9923771} frequently taints measurement data, which can materially impair the accuracy of the traditional CKF algorithm.

To improve this situation the CKF algorithm takes advantage of the fact that information contaminated by non-Gaussian noise does not work well. In recent years, cost functions (learning criteria) based on information theoretic learning (ITL) have received a lot of attention, they have also been widely combined with CKF. Several CKF algorithms incorporating maximum correntropy criterion (MCC) are proposed \cite{LIU2018195,li2020improved,he2020variational}. In addition, the CKF algorithms based on a variant of the MCC are studied \cite{ma2022robust, wang2022robust}. Furthermore, a new robust learning criterion called minimum error entropy (MEE), ITL, performs better than MCC. The CKF algorithms \cite{9646176} based on MEE and mixture MEE is a natural development. 

However, the Gaussian function, in MEE and mixture MEE \cite{HE20221362}, is invariably the best choice for kernel function, further, a more robust generalized MEE (GMEE) learning criterion is proposed \cite{HE2023109188,HE2022}. It is naturally inferred that the GMEE criterion is expected to improve the performance of the existing CKFs, furthermore, the existing GMEE criterion aims to minimize the difference in error, which may lead to errors that do not converge to near zero. These two points constitute the main motivation for this letter.

In this letter, a modified GMEE criterion with a fiducial point (GMEEFP) is proposed to ensure that the error converges to around zero. A new CKF method based on the proposed GMEEFP criterion is developed. A few simulations are implemented to prove the algorithm's feasibility.

\section{Problem formulation}
A nonlinear dynamic system is presented
\begin{align}
\left\{ \begin{gathered}
  {{\boldsymbol{x}}_k} = {\boldsymbol{f}}\left( {{{\boldsymbol{x}}_{k - 1}}} \right) + {{\boldsymbol{q}}_{k - 1}}, \hfill \\
  {{\boldsymbol{y}}_k} = {\boldsymbol{h}}\left( {{{\boldsymbol{x}}_k}} \right) + {{\boldsymbol{r}}_k}. \hfill \\ 
\end{gathered}  \right.
\end{align}
Here ${{\boldsymbol{x}}_k} \in {\mathbb{R}^{n \times 1}}$ represents the state vector at moment $k$, ${{\boldsymbol{y}}_k} \in {\mathbb{R}^{m \times 1}}$ stands for the measurement vector; the state transfer and measurement function are ${\boldsymbol{f}}\left(  \cdot  \right)$ and ${\boldsymbol{h}}\left(  \cdot  \right)$; ${{\boldsymbol{q}}_{k - 1}}$ and ${{\boldsymbol{r}}_k}$ are zero-mean process and measurement noises with covariance matrix ${{\boldsymbol{Q}}_{k - 1}}$ and ${{\boldsymbol{R}}_k}$. The traditional CKF is a classical algorithm that uses observed information to derive an estimation of ${{\boldsymbol{x}}_k}$. Prediction and update are the main steps of the class conventional CKF method.
\subsubsection{Prediction Step}
generate cubature points ${{\boldsymbol{\xi }}_{i;k - 1|k - 1}}$ using $
{{\boldsymbol{\xi }}_{i;k - 1|k - 1}} = {{\boldsymbol{S}}_{k - 1|k - 1}}{{\boldsymbol{\varphi }}_i} + {{\boldsymbol{\hat x}}_{k - 1|k - 1}}$.
Here ${{\boldsymbol{\varphi }}_i}$ is set as ${{\boldsymbol{\varphi }}_i} = \sqrt n {{\boldsymbol{a}}_i}$ for $i = 1,2, \cdots ,n$ and ${{\boldsymbol{\varphi }}_i} =  - \sqrt n {{\boldsymbol{a}}_i}$ for $i = n + 1, \cdots ,2n$, and ${{\boldsymbol{a}}_i}$ represents the unit vector; ${{\boldsymbol{S}}_{k - 1|k - 1}}$ can be obtained by the Cholesky decomposition of ${{\boldsymbol{P}}_{k - 1|k - 1}}$.

Perform propagation calculations for ${{{\boldsymbol{\xi }}_{i;k - 1|k - 1}}}$ using
\begin{align}
{{\boldsymbol{X}}_{i;k|k - 1}} = {\boldsymbol{f}}\left( {{{\boldsymbol{\xi }}_{i;k - 1|k - 1}}} \right),\left( {i = 1,2, \cdots 2n} \right).
\end{align}
Calculate ${{\boldsymbol{\hat x}}_{k|k - 1}}$ and ${{\boldsymbol{P}}_{xx;k|k - 1}}$ by fusing all ${{{\boldsymbol{X}}_{i;k|k - 1}}}$ with weight ${1 \mathord{\left/
 {\vphantom {1 {2n}}} \right.
 \kern-\nulldelimiterspace} {2n}}$
\begin{align}
\left\{ \begin{gathered}
  {{{\boldsymbol{\hat x}}}_{k|k - 1}} = \frac{1}{{2n}}\sum\limits_{i = 1}^{2n} {{{\boldsymbol{X}}_{i;k|k - 1}}} , \hfill \\
  {{\boldsymbol{P}}_{xx;k|k - 1}} = \frac{1}{{2n}}\sum\limits_{i = 1}^{2n} {{{{\boldsymbol{\hat X}}}_{i;k|k - 1}}{\boldsymbol{\hat X}}_{i;k|k - 1}^T + {{\boldsymbol{Q}}_{k - 1}}} , \hfill \\ 
\end{gathered}  \right.
\end{align}
where ${{{\boldsymbol{\hat X}}}_{i;k|k - 1}} = {{\boldsymbol{X}}_{i;k|k - 1}} - {{{\boldsymbol{\hat x}}}_{k|k - 1}}$, ${\left(  \cdot  \right)^T}$ is the transpose operation of a matrix.
\subsubsection{Update Step}
determine cubature points ${{\boldsymbol{\xi }}_{i;k|k - 1}}$ utilizing ${{\boldsymbol{\xi }}_{i;k|k - 1}} = {{\boldsymbol{S}}_{k|k - 1}}{{\boldsymbol{\varphi }}_i} + {{\boldsymbol{\hat x}}_{k|k - 1}}$,
where ${{\boldsymbol{S}}_{k|k - 1}}$ can be obtained utilizing the Cholesky decomposition of ${{\boldsymbol{P}}_{xx;k|k - 1}}$. Then, calculated ${{{\boldsymbol{\xi }}_{i;k|k - 1}}}$ using
\begin{align}
{{\boldsymbol{\gamma }}_{i;k}} = {\boldsymbol{h}}\left( {{{\boldsymbol{\xi }}_{i;k|k - 1}}} \right),\left( {i = 1,2, \cdots ,2n} \right).
\end{align}

Then the predicted measurement ${{{\boldsymbol{\hat y}}}_{k|k - 1}}$ vector, matrices ${{\boldsymbol{P}}_{yy;k|k - 1}}$ and ${{\boldsymbol{P}}_{xy;k|k - 1}}$ can be obtained by using
\begin{align}
\left\{ \begin{gathered}
  {{{\boldsymbol{\hat y}}}_{k|k - 1}} = \frac{1}{{2n}}\sum\limits_{i = 1}^{2n} {{{\boldsymbol{\gamma }}_{i;k}}} , \hfill \\
  {{\boldsymbol{P}}_{yy;k|k - 1}} = \frac{1}{{2n}}\sum\limits_{i = 1}^{2n} {{{{\boldsymbol{\hat \gamma }}}_{i;k}}{\boldsymbol{\hat \gamma }}_{i;k}^T + {{\boldsymbol{R}}_k}} , \hfill \\
  {{\boldsymbol{P}}_{xy;k|k - 1}} = \frac{1}{{2n}}\sum\limits_{i = 1}^{2n} {{{{\boldsymbol{\hat X}}}_{i;k|k - 1}}{\boldsymbol{\hat \gamma }}_{i;k}^T} , \hfill \\ 
\end{gathered}  \right.
\end{align}
where ${{{\boldsymbol{\hat \gamma }}}_{i;k}} = {{\boldsymbol{\gamma }}_{i;k}} - {{{\boldsymbol{\hat y}}}_{k|k - 1}}$. 

Calculate the posterior state vector ${{\boldsymbol{\hat x}}_{k|k}}$ and covariance ${{\boldsymbol{P}}_{k|k}}$ utilizing
\begin{align}
\left\{ \begin{gathered}
  {{{\boldsymbol{\hat x}}}_{k|k}} = {{{\boldsymbol{\hat x}}}_{k|k - 1}} + {{\boldsymbol{K}}_k}\left( {{{\boldsymbol{y}}_k} - {{{\boldsymbol{\hat y}}}_{k|k - 1}}} \right), \hfill \\
  {{\boldsymbol{P}}_{k|k}} = {{\boldsymbol{P}}_{xx;k|k - 1}} - {{\boldsymbol{K}}_k}{{\boldsymbol{P}}_{yy;k|k - 1}}{\boldsymbol{K}}_k^T \hfill \\ 
\end{gathered}  \right.
\end{align}
with the Kalman gain ${{\boldsymbol{K}}_k} = {{\boldsymbol{P}}_{xy;k|k - 1}}{\boldsymbol{P}}_{yy;k|k - 1}^{ - 1}$.

However, due to the impulse disturbance, outliers, or other factors, the distributions of ${{\boldsymbol{r}}_k}$ are generally no longer Gaussian, and reveal the heavy-tail properties. Such non-Gaussian distributions will degrade the performance of the existing KF algorithms since they are initially devised under Gaussian assumptions. To deal with the performance degradation, in this work, a robust KF algorithm is developed to estimate the state ${{\boldsymbol{x}}_k}$ utilizing the information ${{\boldsymbol{y}}_k}$ contaminated by non-Gaussian noise. Specifically, a GMEEFP criterion is developed, and it is combined with the cubature KF filter to dampen the negative effect of the non-Gaussian noises.

\section{CKF based on GMEE with fiducial point}\label{sec:guidelines}
This part develops a modified GMEE criterion with fiducial point, and the cubature KF combined with the proposed criterion is presented.
\subsection{The GMEE with fiducial point}
The information potential (IP) ${\hat V_{\alpha ,\beta }}\left( {{\boldsymbol{X}},{\boldsymbol{Y}}} \right)$ of GMEE criterion \cite{HE2023109188,HE2022} is presented in \eqref{valphabeta}
\begin{align}\label{valphabeta}
{{\hat V}_{\alpha ,\beta }}\left( {{\boldsymbol{X}},{\boldsymbol{Y}}} \right) = {{\hat V}_{\alpha ,\beta }}\left( {\boldsymbol{e}} \right) = \frac{1}{{{N^2}}}\sum\limits_{i = 1}^N {\sum\limits_{j = 1}^N {{G_{\alpha ,\beta }}\left( {{e_i} - {e_j}} \right)} } ,
\end{align}
where ${\boldsymbol{X}}$ and ${\boldsymbol{Y}}$ denote random vectors, parameter $\alpha  > 0$ and $\beta > 0$ are shape parameter and scale parameter; $N$ stands for the number of error in ${\boldsymbol{e}} = [{e_1},{e_2}, \cdots ,{e_N}]$, ${G_{\alpha ,\beta }}\left( e \right) = \left[ {{\alpha  \mathord{\left/
 {\vphantom {\alpha  {2\beta \Gamma \left( {1/\alpha } \right)}}} \right.
 \kern-\nulldelimiterspace} {2\beta \Gamma \left( {1/\alpha } \right)}}} \right]\exp \left( { - {{{{\left| e \right|}^\alpha }} \mathord{\left/
 {\vphantom {{{{\left| e \right|}^\alpha }} {{\beta ^\alpha }}}} \right.
 \kern-\nulldelimiterspace} {{\beta ^\alpha }}}} \right)$ represents the generalized Gaussian density \cite{HE2023108787}.

From \eqref{valphabeta}, one can obtain that the function of \eqref{valphabeta} is to be able to minimize the disparities among errors, which may result in the errors not converging to around 0, for example, each error is large, but the difference among them is small. To address this shortcoming of the GMEE criterion, we construct a modified error vector ${{\boldsymbol{e}}_m} = \left[ {{e_0},{\boldsymbol{e}}} \right]$, where ${e_0} = 0$ represents a constant error that provides a reliable datum for all errors. Considering the fiducial point, the IP ${{\hat V}_{\alpha ,\beta }}\left( {\boldsymbol{e}} \right)$ of the generalized error entropy can be rewritten as
\begin{align}
\begin{gathered}
  {{\hat V}_{\alpha ,\beta }}\left( {{{\boldsymbol{e}}_m}} \right) = \frac{1}{{{{\left( {N + 1} \right)}^2}}}\sum\limits_{i = 0}^N {\sum\limits_{j = 0}^N {{G_{\alpha ,\beta }}\left( {{e_i} - {e_j}} \right)} }  \hfill \\
   = \frac{1}{{{{\left( {N + 1} \right)}^2}}}\left[ \begin{gathered}
  2\sum\limits_{i = 1}^N {{G_{\alpha ,\beta }}\left( {{e_i}} \right)}  + {G_{\alpha ,\beta }}\left( 0 \right) +  \hfill \\
  \sum\limits_{i = 1}^N {\sum\limits_{j = 1}^N {{G_{\alpha ,\beta }}\left( {{e_i} - {e_j}} \right)}  + {G_{\alpha ,\beta }}\left( 0 \right)}  \hfill \\ 
\end{gathered}  \right]. \hfill \\ 
\end{gathered} 
\end{align}

Minimizing the generalized error entropy with fiducial point implies maximizing the IP, and constants ${{G_{\alpha ,\beta }}\left( 0 \right)}$ and ${1 \mathord{\left/
 {\vphantom {1 {{{\left( {N + 1} \right)}^2}}}} \right.
 \kern-\nulldelimiterspace} {{{\left( {N + 1} \right)}^2}}}$ do not affect the result of maximizing the IP. The leraning criterion is called GMEEFP criterion. Therefore, constants ${{G_{\alpha ,\beta }}\left( 0 \right)}$ and ${1 \mathord{\left/
 {\vphantom {1 {{{\left( {N + 1} \right)}^2}}}} \right.
 \kern-\nulldelimiterspace} {{{\left( {N + 1} \right)}^2}}}$ are ignored, and we can obtain
\begin{align}\label{JEIEJjinaNN}
J = 2\sum\limits_{i = 1}^N {{G_{{\alpha _1},{\beta _1}}}\left( {{e_i}} \right)}  + \sum\limits_{i = 1}^N {\sum\limits_{j = 1}^N {{G_{{\alpha _2},{\beta _2}}}\left( {{e_i} - {e_j}} \right)} } .
\end{align}
From \eqref{JEIEJjinaNN}, it can be derived that the new IP is a linear combination of the generalized maximum correntropy and the GMEE IP. In order to balance the ratio of these two IPs, \eqref{JEIEJjinaNN} can be written as
\begin{align}\label{eijejb22asi}
J = \lambda \sum\limits_{i = 1}^N {{G_{{\alpha _1},{\beta _1}}}\left( {{e_i}} \right)}  + \left( {1 - \lambda } \right)\sum\limits_{i = 1}^N {\sum\limits_{j = 1}^N {{G_{{\alpha _2},{\beta _2}}}\left( {{e_i} - {e_j}} \right)} } ,
\end{align}
where $\lambda  \in \left[ {0,1} \right]$ is equilibrium factor. The best result can be reached using the GMEEFP criterion when the errors are forced to decrease to zero. From \eqref{eijejb22asi}, one can obtain that the GMEEFP criterion combines the features of the generalized MCC and GMEE, where the GMEE term minimizes the is able to minimize the difference among errors, the MCC serves to fix all errors around 0, the scaling factor is able to balance the percentage between GMEE and GMCC.

\begin{remark}
When ${\alpha _1} = {\alpha _2} = 2$, \eqref{eijejb22asi} reduces to a linear combination of the MCC and the MEE IP, which means the MEE with fiducial point \cite{9646176} is a special case of GMEEFP.
\end{remark}
\begin{remark}
When $\lambda  = 1$, the GMEEFP criterion reduces to generalized maximum correntropy; when $\lambda  = 0$, the GMEEFP criterion reduces to GMEE criterion. It is clear that the generalized maximum correntropy and GMEE criteria are special cases of the GMEEFP criterion.
\end{remark}

\subsection{The proposed Cubature Kalman filter}
In the regression-based KF solution, the measurement equation and filter update are reformulated as a regression problem \cite{karlgaard2015nonlinear}, therefore, the measurement function and state prediction error are combined to create a regression model with the form of
\begin{align}\label{rkskj1jiaxkh}
\left[ {\begin{array}{*{20}{c}}
  {{{{\boldsymbol{\hat x}}}_{k|k - 1}}} \\ 
  {{{\boldsymbol{y}}_k}} 
\end{array}} \right] = \left[ {\begin{array}{*{20}{c}}
  {{{\boldsymbol{x}}_k}} \\ 
  {{\boldsymbol{h}}\left( {{{\boldsymbol{x}}_k}} \right)} 
\end{array}} \right] + \left[ {\begin{array}{*{20}{c}}
  { - {{\boldsymbol{\varepsilon }}_{k|k - 1}}} \\ 
  {{{\boldsymbol{r}}_k}} 
\end{array}} \right],
\end{align}
where ${{\boldsymbol{\varepsilon }}_{k|k - 1}} = {{\boldsymbol{x}}_k} - {{{\boldsymbol{\hat x}}}_{k|k - 1}}$ represents the prediction error. The hidden state ${{{\boldsymbol{x}}_k}}$ is challenging to extract from the nonlinear measurement equation. A linearized measurement function can be derived by using the statistical linearization in \cite{8338134} as shown below:
\begin{align}\label{uyluuvkerk}
{{\boldsymbol{y}}_k} = {{\boldsymbol{\hat y}}_{k|k - 1}} + {{\boldsymbol{H}}_k}{{\boldsymbol{\varepsilon }}_{k|k - 1}} + {{\boldsymbol{r}}_k} + {{\boldsymbol{v}}_k},
\end{align}
where the linearized matrix ${{\boldsymbol{H}}_k}$ is obtained using ${{\boldsymbol{H}}_k} = {\left( {{\boldsymbol{P}}_{xx;k|k - 1}^{ - 1}{\boldsymbol{P}}_{xy;k|k - 1}^{ - 1}} \right)^T}$.
Combining \eqref{uyluuvkerk}, \eqref{rkskj1jiaxkh} can be further wirtten as
\begin{align}\label{mkxkIhkzzd}
\left[ {\begin{array}{*{20}{c}}
  {{{{\boldsymbol{\hat x}}}_{k|k - 1}}} \\ 
  {{{\boldsymbol{y}}_k} - {{{\boldsymbol{\hat y}}}_{k|k - 1}} + {{\boldsymbol{H}}_k}{{{\boldsymbol{\hat x}}}_{k|k - 1}}} 
\end{array}} \right] = \left[ {\begin{array}{*{20}{c}}
  {{{\boldsymbol{I}}_n}} \\ 
  {{{\boldsymbol{H}}_k}} 
\end{array}} \right]{{\boldsymbol{x}}_k} + {{\boldsymbol{\mu }}_k}
\end{align}
with
\begin{align}
{{\boldsymbol{\mu }}_k} = \left[ {\begin{array}{*{20}{c}}
  { - {{\boldsymbol{\varepsilon }}_{k|k - 1}}} \\ 
  {{{\boldsymbol{r}}_k} + {{\boldsymbol{v}}_k}} 
\end{array}} \right].
\end{align}
where ${{{\boldsymbol{I}}_n}}$ stands for an identity matrix. The covariance of augmented error ${{\boldsymbol{\mu }}_k}$ is calculated using
\begin{align}
\begin{gathered}
  E\left[ {{{\boldsymbol{\mu }}_k}{\boldsymbol{\mu }}_k^T} \right] = {{\boldsymbol{\Theta }}_k}{\boldsymbol{\Theta }}_k^T \hfill \\
   = \left[ {\begin{array}{*{20}{c}}
  {{{\boldsymbol{\Theta }}_{p;k|k - 1}}{\boldsymbol{\Theta }}_{p;k|k - 1}^T}&{\boldsymbol{0}} \\ 
  {\boldsymbol{0}}&{{{\boldsymbol{\Theta }}_{r;k}}{\boldsymbol{\Theta }}_{r;k}^T} 
\end{array}} \right], \hfill \\ 
\end{gathered} 
\end{align}
where ${{\boldsymbol{\Theta }}_k}$, ${{{\boldsymbol{\Theta }}_{p;k|k - 1}}}$, and ${{{\boldsymbol{\Theta }}_{r;k}}}$ can be achieved using the Cholesky decomposition of $E\left[ {{{\boldsymbol{\mu }}_k}{\boldsymbol{\mu }}_k^T} \right]$, ${{\boldsymbol{P}}_{xx;k|k - 1}}$, and ${{\boldsymbol{P}}_{yy;k|k - 1}} + {\boldsymbol{P}}_{xy;k|k - 1}^T{\boldsymbol{P}}_{xx;k|k - 1}^{ - 1}{{\boldsymbol{P}}_{xy;k|k - 1}}$, respectively. 

We can obtained \eqref{dkwkxedk} by multiplying both sides of \eqref{mkxkIhkzzd}
\begin{align}\label{dkwkxedk}
{{\boldsymbol{d}}_k} = {{\boldsymbol{W}}_k}{{\boldsymbol{x}}_k} + {{\boldsymbol{e}}_k}
\end{align}
with
\begin{align}
{{\boldsymbol{d}}_k} = {\boldsymbol{\Theta }}_k^{ - 1}\left[ {\begin{array}{*{20}{c}}
  {{{{\boldsymbol{\hat x}}}_{k|k - 1}}} \\ 
  {{{\boldsymbol{y}}_k} - {{{\boldsymbol{\hat y}}}_{k|k - 1}} + {{\boldsymbol{H}}_k}{{{\boldsymbol{\hat x}}}_{k|k - 1}}} 
\end{array}} \right],
\end{align}
\begin{align}
{{\boldsymbol{W}}_k} = {\boldsymbol{\Theta }}_k^{ - 1}\left[ {\begin{array}{*{20}{c}}
  {{{\boldsymbol{I}}_n}} \\ 
  {{{\boldsymbol{H}}_k}} 
\end{array}} \right],
\end{align}
and
\begin{align}
{{\boldsymbol{e}}_k} = {\boldsymbol{\Theta }}_k^{ - 1}\left[ {\begin{array}{*{20}{c}}
  { - {{\boldsymbol{\varepsilon }}_{k|k - 1}}} \\ 
  {{{\boldsymbol{r}}_k} + {{\boldsymbol{v}}_k}} 
\end{array}} \right].
\end{align}

According to the proposed GMEEFP criterion, the following is an expression for the cost function:
\begin{align}\label{cosgmeefp}
\begin{gathered}
  {J_{GMEEFP}} = \lambda \sum\limits_{i = 1}^N {{G_{{\alpha _1},{\beta _1}}}\left( {{e_i}} \right)}  +  \hfill \\
  \left( {1 - \lambda } \right)\sum\limits_{i = 1}^N {\sum\limits_{j = 1}^N {{G_{{\alpha _2},{\beta _2}}}\left( {{e_i} - {e_j}} \right)} } , \hfill \\ 
\end{gathered} 
\end{align}
where ${e_i} = {d_{i;k}} - {{\boldsymbol{w}}_{i;k}}{{\boldsymbol{x}}_k}$ and ${d_{i;k}}$ represent the $i$th element of ${{\boldsymbol{e}}_k}$ and ${{\boldsymbol{d}}_k}$ respectively; ${{\boldsymbol{w}}_{i;k}}$ represents the $i$th row of ${{\boldsymbol{W}}_k}$, and $N = m + n$. The optimal estimate of the system state can be achieved by calculating ${{{\boldsymbol{\hat x}}}_k} = \arg \mathop {\max }\limits_{{{\boldsymbol{x}}_k}} {J_{GMEEFP}}\left( {{{\boldsymbol{x}}_k}} \right)$.

Taking the derivative of the \eqref{cosgmeefp} on ${{\boldsymbol{x}}_k}$, and we can obtain 
\begin{align}\label{daoshu}
\frac{{\partial {J_{GMEEFP}}}}{{\partial {{\boldsymbol{x}}_k}}} = {\boldsymbol{W}}_k^{\text{T}}{{\boldsymbol{\Lambda }}_k}{{\boldsymbol{d}}_k} - {\boldsymbol{W}}_k^T{{\boldsymbol{\Lambda }}_k}{{\boldsymbol{W}}_k}{{\boldsymbol{x}}_k}
\end{align}
with
\begin{align}
\left\{ \begin{gathered}
  {{\boldsymbol{\Lambda }}_k} = {\lambda _1}{{\boldsymbol{\Pi }}_k} + {\lambda _2}\left( {{{\boldsymbol{\Psi }}_k} - {{\boldsymbol{\Phi }}_k}} \right), \hfill \\
  {\left[ {{{\boldsymbol{\Psi }}_k}} \right]_{ij}} = \left\{ \begin{gathered}
  \sum\limits_{j = 1}^N {{G_{{\alpha _2},{\beta _2}}}\left( {{e_{i;k}} - {e_{j;k}}} \right){{\left| {{e_{i;k}} - {e_{j;k}}} \right|}^{\alpha  - 2}},i = j,}  \hfill \\
  0,i \ne j, \hfill \\ 
\end{gathered}  \right. \hfill \\
  {\left[ {{{\boldsymbol{\Phi }}_k}} \right]_{ij}} = {G_{{\alpha _2},{\beta _2}}}\left( {{e_{j;k}} - {e_{i;k}}} \right){\left| {{e_{j;k}} - {e_{i;k}}} \right|^{\alpha  - 2}}, \hfill \\
  {\left[ {{{\boldsymbol{\Pi }}_k}} \right]_{ij}} = \left\{ \begin{gathered}
  {G_{{\alpha _1},{\beta _1}}}\left( {{e_i}} \right){\left| {{e_i}} \right|^{{\alpha _1} - 2}},i = j, \hfill \\
  0,i \ne j, \hfill \\ 
\end{gathered}  \right. \hfill \\
  {\lambda _1} = \lambda \left( {{{{\alpha _1}} \mathord{\left/
 {\vphantom {{{\alpha _1}} {{\beta _1}^{{\alpha _1}}}}} \right.
 \kern-\nulldelimiterspace} {{\beta _1}^{{\alpha _1}}}}} \right), \hfill \\
  {\lambda _2} = \left( {1 - \lambda } \right)\left( {{{2{\alpha _2}} \mathord{\left/
 {\vphantom {{2{\alpha _2}} {{\beta _2}^{{\alpha _2}}}}} \right.
 \kern-\nulldelimiterspace} {{\beta _2}^{{\alpha _2}}}}} \right). \hfill \\ 
\end{gathered}  \right.
\end{align}
The derivative of \eqref{daoshu} is set to zero. Similar to the derivation in \cite{HE2022}, we can obtain 
\begin{align}\label{dkokwktj1}
{{\boldsymbol{x}}_k}{\text{ = }}{\left( {{\boldsymbol{W}}_k^T{{\boldsymbol{\Omega }}_k}{{\boldsymbol{W}}_k}} \right)^{ - 1}}{\boldsymbol{W}}_k^{\text{T}}{{\boldsymbol{\Omega }}_k}{{\boldsymbol{d}}_k},
\end{align}
where ${{\boldsymbol{\Omega }}_k} = {\lambda _1}{{\boldsymbol{\Pi }}_k} + {\lambda _2}\left( {{\boldsymbol{\Psi }}_k^T{{\boldsymbol{\Psi }}_k} + {\boldsymbol{\Phi }}_k^T{{\boldsymbol{\Phi }}_k}} \right)$. It is clear that \eqref{dkokwktj1} is a function on ${{\boldsymbol{x}}_k}$. Hence, a fixed point iterative (FPI) equation is as follows:
\begin{align}
{{{\boldsymbol{\hat x}}}_{k;t + 1}}{\text{ = }}{\boldsymbol{f}}\left( {{{{\boldsymbol{\hat x}}}_{k;t}}} \right) = {\left( {{\boldsymbol{W}}_k^T{{\boldsymbol{\Omega }}_{k;t}}{{\boldsymbol{W}}_k}} \right)^{ - 1}}{\boldsymbol{W}}_k^{\text{T}}{{\boldsymbol{\Omega }}_{k;t}}{{\boldsymbol{d}}_k},
\end{align}
where $t$ is the number of the FPI, and the initial value of the FPI is ${{{\boldsymbol{\hat x}}}_{k;0}} = {{{\boldsymbol{\hat x}}}_{k|k - 1}}$.

The matrix ${{\boldsymbol{\Omega }}_{k;t}}$ can also be expressed as follows:
\begin{align}\label{28foltkyxf}
{{\boldsymbol{\Omega }}_{k;t}}{\text{ = }}\left[ {\begin{array}{*{20}{c}}
  {{{\boldsymbol{\Omega }}_{x;k;t}}}&{{{\boldsymbol{\Omega }}_{yx;k;t}}} \\ 
  {{{\boldsymbol{\Omega }}_{xy;k;t}}}&{{{\boldsymbol{\Omega }}_{y;k;t}}} 
\end{array}} \right]
\end{align}
with
\begin{align}
\left\{ \begin{gathered}
  {{\boldsymbol{\Omega }}_{x;k;t}} \in {\mathbb{R}^{n \times n}},{{\boldsymbol{\Omega }}_{xy;k;t}} \in {\mathbb{R}^{m \times n}}, \hfill \\
  {{\boldsymbol{\Omega }}_{yx;k;t}} \in {\mathbb{R}^{n \times m}},{{\boldsymbol{\Omega }}_{y;k;t}} \in {\mathbb{R}^{m \times m}}. \hfill \\ 
\end{gathered}  \right.
\end{align}
Substituting \eqref{28foltkyxf} into ${\boldsymbol{W}}_k^T{{\boldsymbol{\Omega }}_{k;t}}{{\boldsymbol{W}}_k}$ yields
\begin{align}
\begin{gathered}
  {\boldsymbol{W}}_k^T{{\boldsymbol{\Omega }}_{k;t}}{{\boldsymbol{W}}_k}{\text{ = }}{\boldsymbol{\bar P}}_{k|k - 1;t}^x + {\boldsymbol{H}}_k^T{\boldsymbol{\bar P}}_{k|k - 1;t}^{xy} +  \hfill \\
  \left( {{\boldsymbol{\bar P}}_{k|k - 1;t}^{yx} + {\boldsymbol{H}}_k^T{\boldsymbol{\bar P}}_{k|k - 1;t}^y} \right){\boldsymbol{I}}{{\boldsymbol{H}}_k} \hfill \\ 
\end{gathered} 
\end{align}
with
\begin{align}
\left\{ \begin{gathered}
  {\boldsymbol{\bar P}}_{k|k - 1;t}^x = {\left( {{\boldsymbol{\Theta }}_{p;k|k - 1}^{ - 1}} \right)^T}{{\boldsymbol{\Omega }}_{x;k;t}}{\boldsymbol{\Theta }}_{p;k|k - 1}^{ - 1}, \hfill \\
  {\boldsymbol{\bar P}}_{k|k - 1;t}^{xy} = {\left( {{\boldsymbol{\Theta }}_{r;k}^{ - 1}} \right)^T}{{\boldsymbol{\Omega }}_{xy;k;t}}{\boldsymbol{\Theta }}_{p;k|k - 1}^{ - 1}, \hfill \\
  {\boldsymbol{\bar P}}_{k|k - 1;t}^{yx} = {\left( {{\boldsymbol{\Theta }}_{p;k|k - 1}^{ - 1}} \right)^T}{{\boldsymbol{\Omega }}_{yx;k;t}}{\boldsymbol{\Theta }}_{r;k}^{ - 1}, \hfill \\
  {\boldsymbol{\bar P}}_{k|k - 1;t}^y = {\left( {{\boldsymbol{\Theta }}_{r;k}^{ - 1}} \right)^T}{{\boldsymbol{\Omega }}_{y;k;t}}{\boldsymbol{\Theta }}_{r;k}^{ - 1}. \hfill \\ 
\end{gathered}  \right.
\end{align}
In a similar way, ${\boldsymbol{W}}_k^T{{\boldsymbol{\Omega }}_{k;t}}{{\boldsymbol{d}}_k}$ can be further represented as 
\begin{align}\label{dkouktwt}
\begin{gathered}
  {\boldsymbol{W}}_k^{\text{T}}{{\boldsymbol{\Omega }}_{k;t}}{{\boldsymbol{d}}_k} = {\boldsymbol{\bar P}}_{k|k - 1;t}^x{{{\boldsymbol{\hat x}}}_{k|k - 1}} + {\boldsymbol{H}}_k^T{\boldsymbol{\bar P}}_{k|k - 1;t}^{xy}{{{\boldsymbol{\hat x}}}_{k|k - 1}} \hfill \\
   + {\boldsymbol{\bar P}}_{k|k - 1;t}^{yx}\left( {{{\boldsymbol{y}}_k} - {{{\boldsymbol{\hat y}}}_{k|k - 1}} + {{\boldsymbol{H}}_k}{{{\boldsymbol{\hat x}}}_{k|k - 1}}} \right) +  \hfill \\
  {\boldsymbol{H}}_k^T{\boldsymbol{\bar P}}_{k|k - 1;t}^y\left( {{{\boldsymbol{y}}_k} - {{{\boldsymbol{\hat y}}}_{k|k - 1}} + {{\boldsymbol{H}}_k}{{{\boldsymbol{\hat x}}}_{k|k - 1}}} \right). \hfill \\ 
\end{gathered} 
\end{align}

For calculating ${\boldsymbol{W}}_k^T{{\boldsymbol{\Omega }}_{k;t}}{{\boldsymbol{W}}_k}$, the matrix inversion lemma is employed, and we can obtain
\begin{align}\label{tkjxyphTi}
\begin{gathered}
  {\left( {{\boldsymbol{W}}_k^T{{\boldsymbol{\Omega }}_{k;t}}{{\boldsymbol{W}}_k}} \right)^{ - 1}} = {\left( {{\boldsymbol{\bar P}}_{k|k - 1;t}^x + {\boldsymbol{H}}_k^T{\boldsymbol{\bar P}}_{k|k - 1;t}^{xy}} \right)^{ - 1}} -  \hfill \\
  {\left( {{\boldsymbol{\bar P}}_{k|k - 1;t}^x + {\boldsymbol{H}}_k^T{\boldsymbol{\bar P}}_{k|k - 1;t}^{xy}} \right)^{ - 1}}\left( {{\boldsymbol{\bar P}}_{k|k - 1;t}^{yx} + {\boldsymbol{H}}_k^T{\boldsymbol{\bar P}}_{k|k - 1;t}^y} \right) \hfill \\
   \times {\left[ \begin{gathered}
  {\boldsymbol{I}} + {{\boldsymbol{H}}_k}{\left( {{\boldsymbol{\bar P}}_{k|k - 1;t}^x + {\boldsymbol{H}}_k^T{\boldsymbol{\bar P}}_{k|k - 1;t}^{xy}} \right)^{ - 1}} \hfill \\
   \times \left( {{\boldsymbol{\bar P}}_{k|k - 1;t}^{yx} + {\boldsymbol{H}}_k^T{\boldsymbol{\bar P}}_{k|k - 1;t}^y} \right) \hfill \\ 
\end{gathered}  \right]^{ - 1}} \times  \hfill \\
  {{\boldsymbol{H}}_k}{\left( {{\boldsymbol{\bar P}}_{k|k - 1;t}^x + {\boldsymbol{H}}_k^T{\boldsymbol{\bar P}}_{k|k - 1;t}^{xy}} \right)^{ - 1}}. \hfill \\ 
\end{gathered} 
\end{align}
Substituting \eqref{dkouktwt} and \eqref{tkjxyphTi} into \eqref{dkokwktj1}, and ${{{\boldsymbol{\hat x}}}_{k;t + 1}}$ can be further written as
\begin{align}
{{\boldsymbol{\hat x}}_{k;t + 1}} = {{\boldsymbol{\hat x}}_{k|k - 1}} + {{\boldsymbol{K}}_{k;t}}\left( {{{\boldsymbol{y}}_k} - {{{\boldsymbol{\hat y}}}_{k|k - 1}}} \right)
\end{align}
with
\begin{align}
{{\boldsymbol{K}}_{k;t}} = {\left( {{\boldsymbol{W}}_k^T{{\boldsymbol{\Omega }}_{k;t}}{{\boldsymbol{W}}_k}} \right)^{ - 1}}\left( {{\boldsymbol{\bar P}}_{k|k - 1;t}^{yx} + {\boldsymbol{H}}_k^T{\boldsymbol{\bar P}}_{k|k - 1;t}^y} \right).
\end{align}
If the result satisfies ${{{\rm{||}}{{{\boldsymbol{\hat x}}}_{k;t + 1}} - {{{\boldsymbol{\hat x}}}_{k;t}}{\rm{||}}} \mathord{\left/
 {\vphantom {{{\rm{||}}{{{\boldsymbol{\hat x}}}_{k;t + 1}} - {{{\boldsymbol{\hat x}}}_{k;t}}{\rm{||}}} {{\rm{||}}{{{\boldsymbol{\hat x}}}_{k;t}}{\rm{||}}}}} \right.
 \kern-\nulldelimiterspace} {{\rm{||}}{{{\boldsymbol{\hat x}}}_{k;t}}{\rm{||}}}} \le \tau $, the FPI loops are considered to be convergent, and ${{\boldsymbol{K}}_{k;t}} = {{\boldsymbol{K}}_k}$.
Finally, the posterior covariance matrix can be updated using
\begin{align}
{{\boldsymbol{P}}_{k|k}} = \left( {{\boldsymbol{I}} - {{\boldsymbol{K}}_k}{{\boldsymbol{H}}_k}} \right){{\boldsymbol{P}}_{k|k - 1}}{\left( {{\boldsymbol{I}} - {{\boldsymbol{K}}_k}{{\boldsymbol{H}}_k}} \right)^{\text{T}}} + {{\boldsymbol{K}}_k}{{\boldsymbol{R}}_k}{\boldsymbol{K}}_k^T.
\end{align}

\section{Simulation}
In this part, the efficiency of the GMEEFP-CKF is compared to that of the CKF \cite{4982682}, MCCKF \cite{9129755}, and MEEF-CKF \cite{9646176}. All simulations are averaged over 200 Monte Carlo runs, where 200 samples are used to calculate the mean-square deviation (MSD) that is utilized to evaluate the effectiveness of the proposed method in relation to its competitors. The concept of MSD is defined as $MSD = 10{\log _{10}}{\rm{||}}{{\boldsymbol{x}}_k} - {{\boldsymbol{\hat x}}_{k|k}}{\rm{|}}{{\rm{|}}^2}$, where ${{{\boldsymbol{x}}_k}}$ denotes the real state of the system.

A vehicle tracking model is considered, and a process equation is given as
\begin{align}
{{\boldsymbol{x}}_k} = \left[ {\begin{array}{*{20}{c}}
  {{{\boldsymbol{I}}_2}}&{\Delta T{{\boldsymbol{I}}_2}} \\ 
  {\boldsymbol{0}}&{{{\boldsymbol{I}}_2}} 
\end{array}} \right]{{\boldsymbol{x}}_{k - 1}} + {{\boldsymbol{q}}_{k - 1}},
\end{align}
where ${{{\boldsymbol{I}}_2}}$ denotes the unit matrix and $\Delta T = 0.5s$. State ${{\boldsymbol{x}}_k} = {\left[ {\begin{array}{*{20}{c}}
  {{p_{1;k}}}&{{p_{2;k}}}&{{v_{1;k}}}&{{v_{2;k}}} 
\end{array}} \right]^{\text{T}}}$ contains position ${{p_{1;k}}}$ and velocity ${{v_{1;k}}}$ of the target in the $x$-axis and the position ${{p_{2;k}}}$ and velocity ${{v_{2;k}}}$ in the $y$-axis. 

The measurement equation with the distance and angle of the target is written as:
\begin{align}\label{scenarios_five}
{{\boldsymbol{z}}_k} = \left[ {\begin{array}{*{20}{c}}
  {\sqrt {p_{1;k}^2 + p_{2;k}^2} } \\ 
  {\arctan \frac{{{p_{2;k}}}}{{{p_{1;k}}}}} 
\end{array}} \right] + {{\boldsymbol{r}}_k},
\end{align}
In the numerical simulation, the process noise of the system is set to Gaussian noise $\mathcal{N}\left( {0,0.1} \right)$, and the measurement noise is set to mixed-Gaussian noise \cite{HE20221362} ${[{{\boldsymbol{r}}_k}]_i} \sim {\text{0}}{\text{.96}}\mathcal{N}\left( {0,1} \right) + {\text{0}}{\text{.04}}\mathcal{N}\left( {0,100} \right)$. The initial values of ${{\boldsymbol{\hat x}}_{0|0}}$ and ${{\boldsymbol{P}}_{0|0}}$ are set to
\begin{align}
\left\{ \begin{gathered}
  {{{\boldsymbol{\hat x}}}_{0|0}} \sim \mathcal{N}\left( {{{\boldsymbol{x}}_0},{{\boldsymbol{I}}_n}} \right), \hfill \\
  {{\boldsymbol{P}}_{0|0}} = {{\boldsymbol{I}}_n}, \hfill \\ 
\end{gathered}  \right.
\end{align}
where ${{\boldsymbol{x}}_0} = {\left[ {1,1,10,20} \right]^{\text{T}}}$ is the true state of target.

\begin{figure}[tb]
\centerline{\includegraphics[width=\columnwidth]{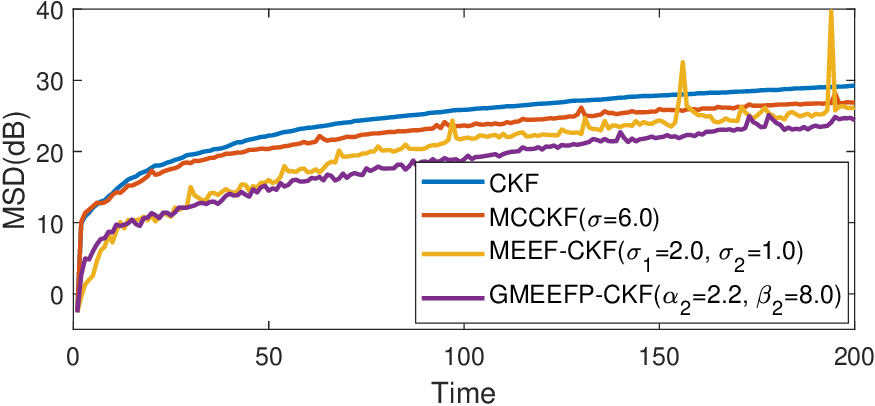}}
\caption{The MSD of different algorithms under mixed-Gaussian noise.}\label{MSD_mixture}
\end{figure}

\begin{figure}[tb]
\centerline{\includegraphics[width=\columnwidth]{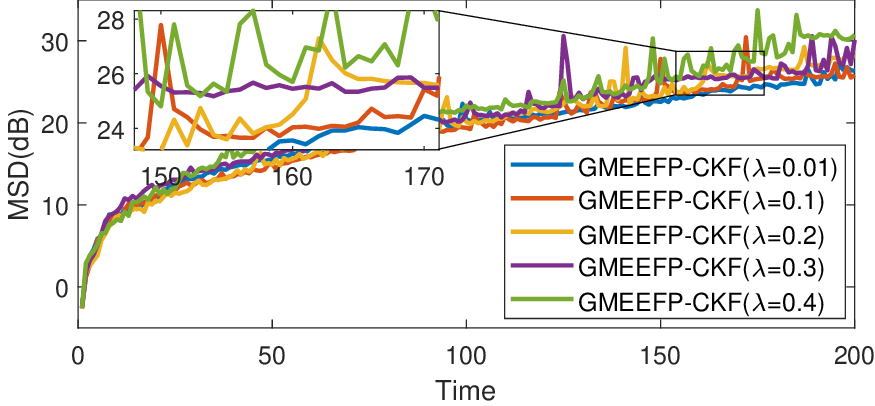}}
\caption{The MSD of the GMEEFP-CKF with different $\lambda$.}\label{MSD_eta}
\end{figure}

Fig. \ref{MSD_mixture} displays the performance of several methods in terms of MSD. Table \ref{MSD_different_parameter} presents the steady MSD of the GMEEFP-CKF method employing different $\alpha$ and $\beta$, and Fig. \ref{MSD_eta} shows the convergence curve of the MSD with different $\lambda$. From these simulation results, one can obtain that 1) the proposed GMEEFP-CKF algorithm outperforms the existing CKF algorithms with mixed-Gaussian noise; 2) When ${\alpha _2} = 2.2$ and ${\beta _2} = 6.0$ the proposed algorithm obtains the optimal performance with mixed-Gaussian noise; the performance of the GMEEFP-CKF method decreases as $\lambda$ increases.

\begin{table}[tb]
\centering
\caption{The MSD (dB) of the GMEEFP-CKF with different parameters.}\label{MSD_different_parameter}
\begin{tabular}{llllll}
\hline
           & ${\beta _2} = 1$ & ${\beta _2} = 2$ & ${\beta _2} = 4$ & ${\beta _2} = 6$ & ${\beta _2} = 8$ \\ \hline
${\alpha _2} = 2.0$   & 27.72   & 25.44   & 23.9    & 25.65   & 26.6    \\
${\alpha _2} = 2.2$ & 33.04   & 28.36   & 23.78   & 22.42   & 23.17   \\
${\alpha _2} = 2.4$ & 33.45   & 27.05   & 23.62   & 23.00   & 23.87   \\
${\alpha _2} = 2.6$ & 31.05   & 24.66   & 23.14   & 24.36   & 24.12   \\
${\alpha _2} = 2.8$ & 30.3    & 26.44   & 25.67   & 24.76   & 25.36   \\
${\alpha _2} = 3.2$ & 29.3    & 25.72   & 25.62   & 27.08   & 29.57   \\
${\alpha _2} = 4.0$ & fail    & 33.66   & 29.11   & 29.31   & 30.53   \\ \hline
\end{tabular}
\end{table}

\section{Conclusion}
In this letter, the GMEEFP criterion is proposed to ensure the error converges to around zero. 
In combination with the GMEEFP criterion, a CKF is derived to reduce the effect of non-Gaussian noise. The suggested technique outperforms existing methods for nonlinear system state estimation with non-Gaussian noise, according to simulation findings.

\bibliographystyle{unsrt}
\bibliography{ref}
\end{document}